\documentclass[letter]{aa}  

\usepackage{graphicx,natbib}
\usepackage{txfonts}
\begin{document}

   \title{Rotation of the asymptotic giant branch star R Doradus}

   \titlerunning{Rotation of the AGB star R Dor}


   \author{W.~H.~T. Vlemmings
          \inst{1}\fnmsep\thanks{wouter.vlemmings@chalmers.se}
          \and
          T. Khouri\inst{1} \and E. De Beck\inst{1} \and
          H. Olofsson\inst{1} \and G. Garc\'{\i}a-Segura\inst{2} \and
          E. Villaver\inst{3}
          A. Baudry\inst{4} \and E.M.L. Humphreys\inst{5} \and
          M. Maercker\inst{1} \and S. Ramstedt\inst{6}
          }

   \institute{Department of Space, Earth and Environment, Chalmers University of Technology, Onsala Space Observatory, 439 92 Onsala, Sweden
       \and
        Instituto de Astronom\'{\i}a, Universidad Nacional Aut\'onoma de M\'exico, Km. 107 Carr. Tijuana-Ensenada, 22860, Ensenada, B. C., Mexico
          \and
          Departamento de F\'{\i}sica Te\'orica, Universidad Aut\'onoma de Madrid, Cantoblanco 28049  Madrid, Spain
  \and
          Laboratoire d'astrophysique de Bordeaux, Univ. Bordeaux, CNRS, B18N, all\'ee Geoffroy Saint-Hilaire, F-33615 Pessac, France
  \and 
         European Southern Observatory, Karl-Schwarzschild-Str. 2,
         85748 Garching, Germany
   \and
   Department of Physics and Astronomy, Uppsala University, Box 516, 751 20, Uppsala, Sweden}

   \date{12-Apr-2018}

  \abstract{ High resolution observations of the extended atmospheres
     of asymptotic giant branch (AGB) stars can now directly confront
     the theories that describe stellar mass loss. Using Atacama Large
     Millimeter/submillimeter Array (ALMA) high angular resolution
     ($30\times42$~mas) observations we have, for the first time,
     resolved stellar rotation of an AGB star, R~Dor. We measure an
     angular rotation velocity of
     $\omega_R\sin{i}=(3.5\pm0.3)\times10^{-9}$~rad~s$^{-1}$ which indicates a rotational
     velocity of $|\upsilon_{\rm
       rot}\sin{i}|=1.0\pm0.1$~km~s$^{-1}$  at
     the stellar surface ($R_*=31.2$~mas at $214$~GHz). The rotation
     axis projected on the plane of the sky has a position angle
     $\Phi=7\pm6^\circ$. We find that the rotation of R Dor is two
     orders of magnitude faster than expected for a solitary AGB star
     that will have lost most of its angular momentum. Its rotational
     velocity is consistent with angular momentum transfer from a
     close companion. As a companion has not been directly detected we
     thus suggest R~Dor has a low-mass, close-in, companion. The
     rotational velocity approaches the critical velocity, set by the
     local sound speed in the extended envelope, and is thus expected
     to affect the mass loss characteristics of R~Dor. }


   \maketitle
%

\section{Introduction}

Rotation of thermally pulsing (TP) AGB stars can have significant
effects on their interior structure and evolution. The internal dynamics
is changed by e.g. the transport of angular momentum and turbulent
mixing, and rotation induced instabilities and rotational
mixing can affect the nucleosynthesis and change relative stellar
yields \citep[e.g.][]{Lagarde2012, Piersanti2013}. Differential
rotation can also generate strong magnetic fields
\citep[e.g.][]{Blackman2001}.
Finally, rotation can directly affect the mass
loss by changing the isotropic wind into one with a strong equatorial
component \citep[e.g.][]{Dorfi1996, Ignace1996}.

Stellar evolution models predict surface rotation velocities for
(solitary) low- to intermediate mass TP-AGB stars of only a few 10s of
m~s$^{-1}$. As far as we know, the carbon AGB star V~Hya is the only AGB star for
which fast rotation has been inferred from spectroscopic observations
\citep{Barnbaum1995}. As V~Hya also displays evidence of a
high-velocity collimated outflow, with multiple components, it has
been suggested that V~Hya is a late-AGB star in a binary system
\citep[e.g.][]{Sahai2016}.  Here we present the first direct detection
of rotation of the extended atmosphere of an AGB star,
R~Dor. \object{R Dor} is a nearby \citep[59\,pc;][]{Knapp2003} M-type
AGB star with a slow wind and a low mass-loss rate
\citep[$\upsilon_{\mathrm{exp}}=5.7$\,km\,s$^{-1}$,
$\dot{M}=1-2\times10^{-7}$\,M$_\odot$\,yr$^{-1}$;][]{Ramstedt2014,Maercker2016}. 
Using the oxygen isotopologue ratios Si$^{17}$O/Si$^{18}$O and H$_2^{17}$O/H$_2^{18}$O
as proxies for the isotopic ratio $^{17}$O/$^{18}$O the initial
stellar mass of R Dor is estimated to be $1.3-1.6$\,M$_\odot$
\citep{DeBeck2018,Danilovich2017}.

\begin{figure*}
\centering
\includegraphics[width=13.5cm]{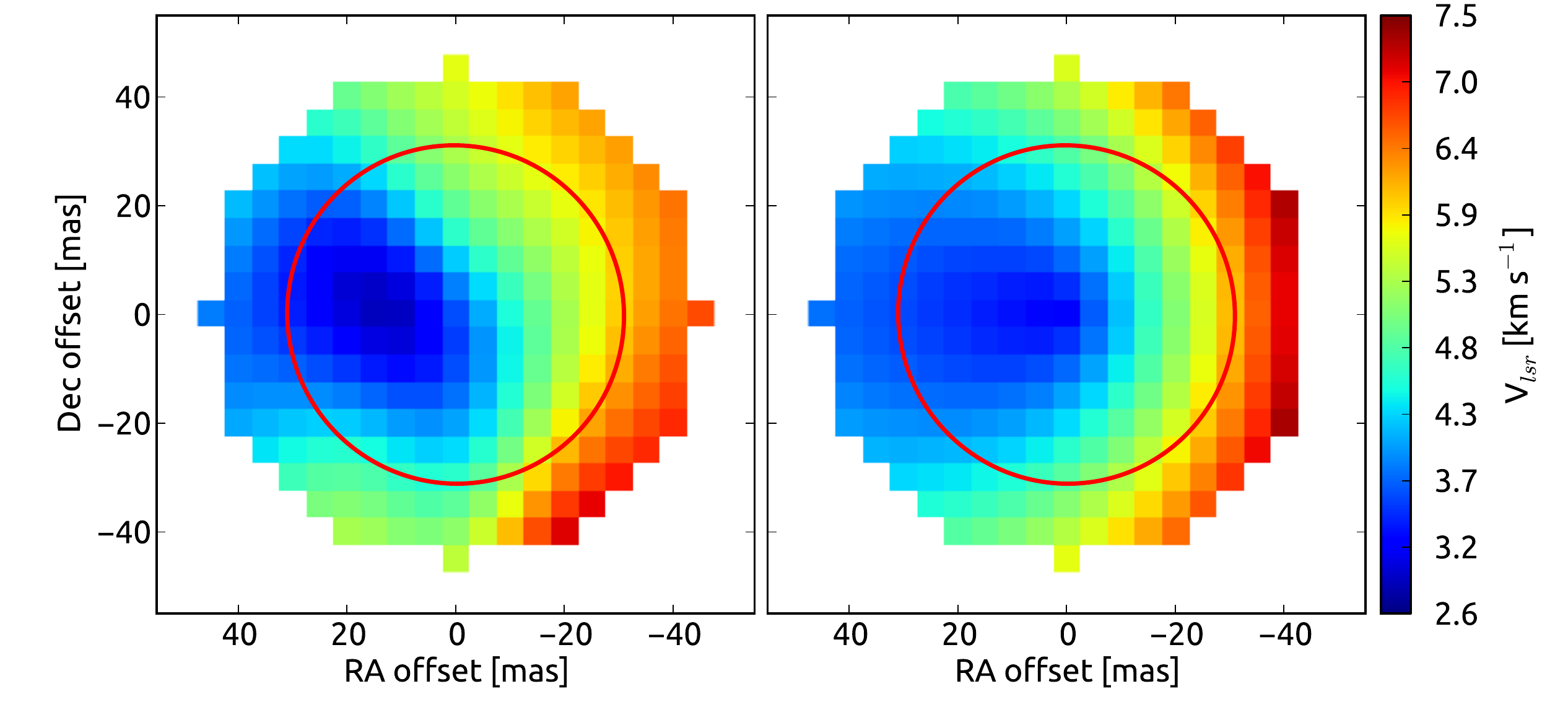}
\caption{{\it (left)}: Center velocity of the SiO $v=3, J=5-4$
emission line as determined from a fit to the observations. In this
and the other panels, the red ellipse indicates the measured size of
the star at $214$~GHz. {\it (right)}: The best fit model of
solid-body rotation (see text).}
 \label{SiOfit}
\end{figure*}

\section{Observations and data reduction}

These observations of R~Dor were performed as part of ALMA project
2016.1.0004.S.
Here we mainly focus on
observations in Band 6, although Band 4 observations
confirm the results.  The observations
took place on 25 September 2017,
and were performed using four spectral windows (spws) of 1920 channels
each. The velocity resolution was $\sim1.4$~km~s$^{-1}$ and the spws were
centered on 213.97, 215.97, 226.29, and 228.21~GHz. Further details of
the calibration and self-calibration will be presented in a forthcoming paper
(Vlemmings et al., in prep.).

The stellar continuum emission of R~Dor is clearly detected and
resolved. Using the uv-fitting procedure decribed in
\citet{Vlemmings2017} we have determined the size of R~Dor
at 214~GHz to be $(62.4\pm0.1)\times(61.8\pm0.1)~{\rm mas}$ at
a position angle of $33\pm6^\circ$ east from north. The continuum
emission was subtracted before the subsequent line imaging.
The line image products were created using Briggs
robust weighting. This resulted in a typical beam size of
$30\times42~{\rm mas}$ and a typical channel rms noise
of $\sim2.5$~mJy~beam$^{-1}$. Since
the data were taken in one of the longest baseline configurations, the
maximum recoverable scale is $\sim0.35\arcsec$. Hence we are mainly
sensitive to strong compact emission.

\section{Results}

\begin{figure*}
\centering
\includegraphics[width=13.5cm]{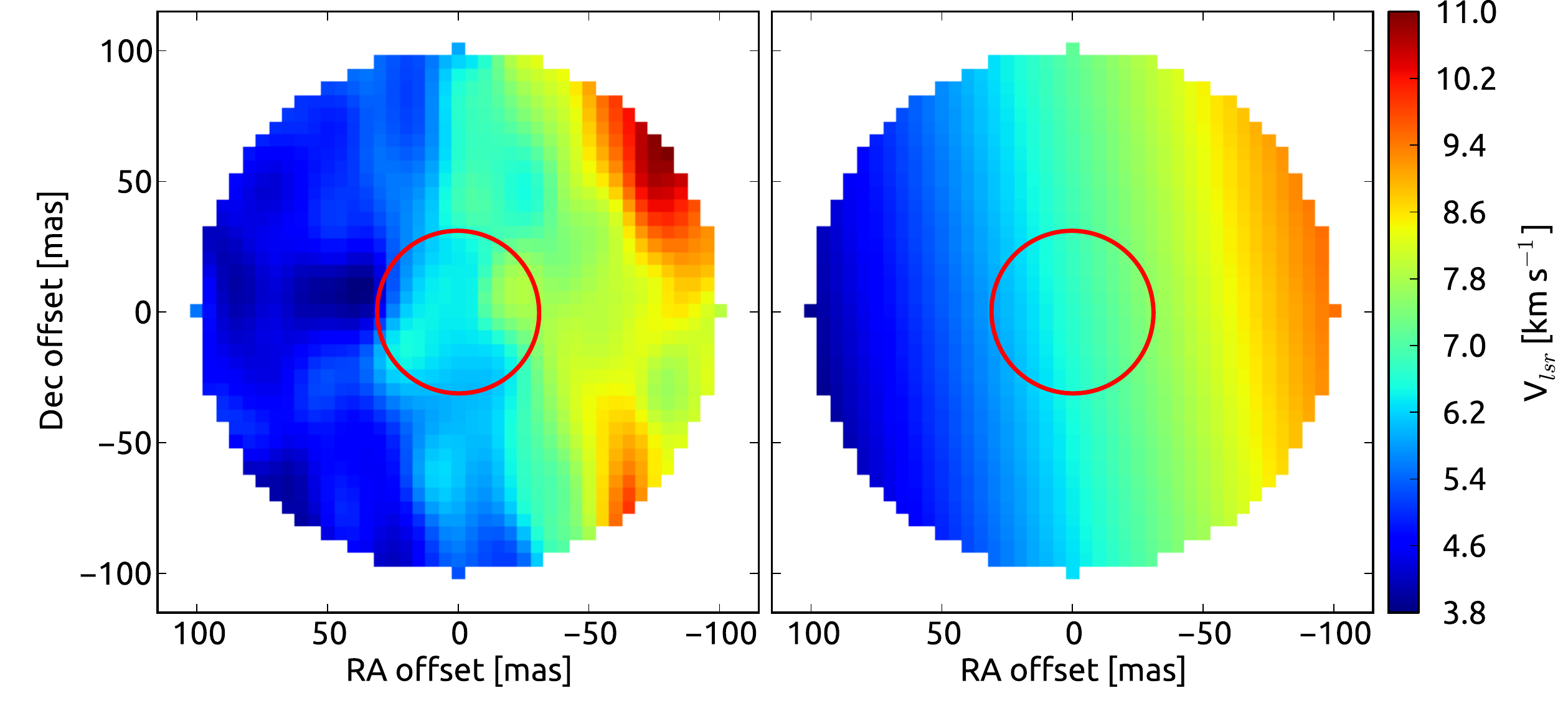}
\caption{As Fig.\ref{SiOfit} for the SO$_2$ $J_{K_{\mathrm{a}},K_{\mathrm{c}}} = 16_{3,13}-16_{2,14}$ line.}
   \label{SO2fit}
\end{figure*}

%


Numerous molecular lines were detected towards R~Dor. Here we
focus mainly on two of the SiO lines and an SO$_2$ line, namely the SiO $v=3,
J=5-4$ and SO$_2$ $J_{K_{\mathrm{a}},K_{\mathrm{c}}} = 16_{3,13}-16_{2,14}$ emission lines and the $^{29}$SiO~$v=1, J=5-4$
absorption line. Hereafter, we denote the lines as SiO $v=3$, SO$_2$
and $^{29}$SiO~$v=1$. The lines are compact and
detected at sufficient signal-to-noise ratio to allow a careful
analysis of their velocity structure within three stellar
radii. Additionally, although some maser action might be present in
the strongest emission peak of the SiO~$v=3$ line, its linewidth
and strength does not indicate a strong masing effect. Any further
discussion on the line emission and absorption characteristics and
excitation is beyond the scope of this Letter.

As the lines in the stellar atmosphere are affected by pulsations and
convective motions, the linewidths are large. We measure widths of
$\sim5-10$~km~s$^{-1}$ for the SO$_2$ line , $\sim10$~km~s$^{-1}$ for the
SiO $v=3$ line, and $\sim15-20$~km~s$^{-1}$ for the $^{29}$SiO~$v=1$ line. In
order to describe the velocity field around the star, we fit the
spectra from each image pixel using the CASA task {\it specfit}. In
Fig.~\ref{lines} we present the line spectra, for the SiO $v=3$,
SO$_2$, and $^{29}$SiO~$v=1$ lines, taken in an aperture
covering the dominant part of the line emission region, and the
resulting fit and residual spectra. We found that a single Lorentzian
profile accurately describes the majority of the spectral lines,
although the strong absorption line of $^{29}$SiO~$v=1$ is better
fitted with two Gaussian components. The fitting task produces images
of intensity, central velocity, full-width-half-maximum (fwhm) line
width and their respective errors. For the SiO $v=3$ and SO$_2$ lines
these, and the fit residuals, are shown in
Figs~\ref{SiOfit},~\ref{SiOres},~\ref{SiOother}, and
~\ref{SO2fit},~\ref{SO2res}, and ~\ref{SO2other} respectively.  The
velocity field of the strongest $^{29}$SiO~$v=1$ component is
presented in Fig.~\ref{29SiOfit} and its residual, error, intensity
and fwhm maps in Figs~\ref{29SiOres} and ~\ref{29SiOother}. Because
this line was seen only in absorption against the star, and hence is
only marginally resolved, we base our main results on the two other
lines. Channel maps for the three lines discussed here are shown in
Appendix B.

The velocity field of the three lines presented here all show a clear
gradient across the entire emitting region from red-shifted to
blue-shifted velocities with respect to the systemic
velocity. Generally, velocity gradients could be caused by an outflow,
an expanding and/or rotating disc or torus, rotation of the entire
emitting structure or a chance alignment of convective cells producing
outward motions. Since we see the velocity gradients out to at least
two $R_*$, we can rule out a chance alignment of convective
cells. At the outflow velocities measured, material would have to
  be coherently ejected from above convective cells for a period of
  more than $2$~years to produce the observed distribution. This is longer
  than typical convective timescales \citep{Freytag2017}. 

Based on the line intensity distributions, we can also rule out an
edge-on torus or disc, since there is no sign of a flattened
distribution across the star. Furthermore, as we observe the
  rotation within $1.1~R_*$, its origin in a disc or torus means that
  this structure would have to extend to the surface of the star,
  without any gap. This is in contrast to a recent tentative claim of
  an edge on disc, with an inner radius of $\sim6$~au, around R~Dor by
  \citet{Homan2018}. They report a velocity gradient, fully consistent
  with our interpretation of solid-body rotation, observed in
  different molecular lines but with significantly lower angular
  resolution ($\sim150$~mas). Because of the lack of resolution, the
  authors are led to suggest a disc. Their tentative interpretation is
  not consistent with the data presented here. Moreover, observations
  of dust scattered light between $2$ and $10~R_*$ do not support the
  idea of an edge on disc \citep[][\& in prep.]{Khouri2016}.

A tilted expanding disc or torus seen almost
face-on could explain the velocities seen in the SO$_2$ and other
extended lines. It would however not produce the pattern of red- to
blue-shifted velocities seen in the SiO lines across the stellar
disc. Similarly, a bipolar outflow would not produce such a pattern across
the stellar disc unless the outflow has an opening
angle close to $180^\circ$ and material is launched almost tangentially to the surface near
the equator. There is no known mechanism that would explain such an
outflow originating at the stellar surface. Consequently we conclude that only
rotation of the star and its envelope can explain the velocity field
observed.

The velocity field was fit, using a $\chi^2-$analysis,
to the velocity field produced by the rotation of a shell with solid-body
rotation. The rotation velocity for solid-body rotation including possible
expansion is given by \citep[e.g.][]{Kervella2018}:
\begin{equation}
\upsilon_{\rm obs} (p) = {{p_y}\over{R_{\rm shell}}}(\upsilon_{\rm rot}\sin{i}) +
(\upsilon_{\rm sys} - \upsilon_{\rm exp}{{(R_{\rm shell}-p)}\over{R_{\rm shell}}}). 
\end{equation}
Here, $\upsilon_{\rm obs}$ is the observed velocity, $p$ is the
projected distance on the sky from the centre of the star and $p_y$ is
the component of $p$ perpendicular to the rotation axis (with position
angle $\Phi$). $R_{\rm shell}$ is the average shell radius of the
emission line, $\upsilon_{\rm rot}$ is the rotation velocity at
$R_{\rm shell}$, and $i$ is the unknown inclination of the rotation
axis. Additional velocity components correspond to the systemic
velocity $\upsilon_{\rm sys}$ of the line and the expansion velocity
$\upsilon_{\rm exp}$ ($\upsilon_{\rm exp}=0$ for $p>R_{\rm
  shell}$). As noted in \citet{Kervella2018}, there is a strong
correlation between $\upsilon_{\rm sys}$ and $\upsilon_{\rm exp}$ if
the observations provide only a few resolution elements across the
emission region, which is the case for our observations. We determine,
for the emission lines, the average shell radius $R_{\rm shell}$ using
{\it immultifit} \citep{MartiVidal2014} by fitting a shell of emission
to each individual channel. We then take $R_{\rm shell}$ to be the
weighted average of the four central emission channels. Finally, the
$\chi^2$ analysis is performed using the maps of pixel-based central
velocity and the associated errors produced using {\it specfit}. We
limit the pixels included to those where the fitted amplitudes are at
least three times the associated fit error.  The results of the fits
are presented in Table~\ref{tab1} and shown, with the measured
velocity fields, in Figs~\ref{SiOfit},~\ref{SO2fit}, and
\ref{29SiOfit}. The velocity error map, used in the $\chi^2$ analysis,
and the residual velocity maps are shown in the same figures. The
amplitude and linewidth maps for the same lines are shown in Appendix
B. Although the approximation of solid-body rotation for each
individual molecular line might not be fully correct, our angular
resolution does not allow us to fit the extra parameters needed for a
more detailed description of the rotation. However, a comparison
between the SiO $v=3$ and SO$_2$ lines that peak at significantly
different radii yield results for the angular rotation velocity that
are consistent within $<1.5\sigma$. We thus conclude that the
extended atmosphere indeed appears to display solid-body rotation out
to at least two $R_*$ and adopt, from a weighted average
  of the SiO $v=3$ and SO$_2$ lines, an angular rotation velocity
of $\omega_R\sin{i}=(3.5 \pm 0.3) \times10^{-9}$~rad~s$^{-1}$ for the
stellar atmosphere of R~Dor. This implies a rotation period $P/\sin{i}
= 57\pm5$~yr. The weighted average position angle of the rotation
  axis $\Phi=7\pm6^\circ$.

\begin{figure}
\centering
\includegraphics[width=8.5cm]{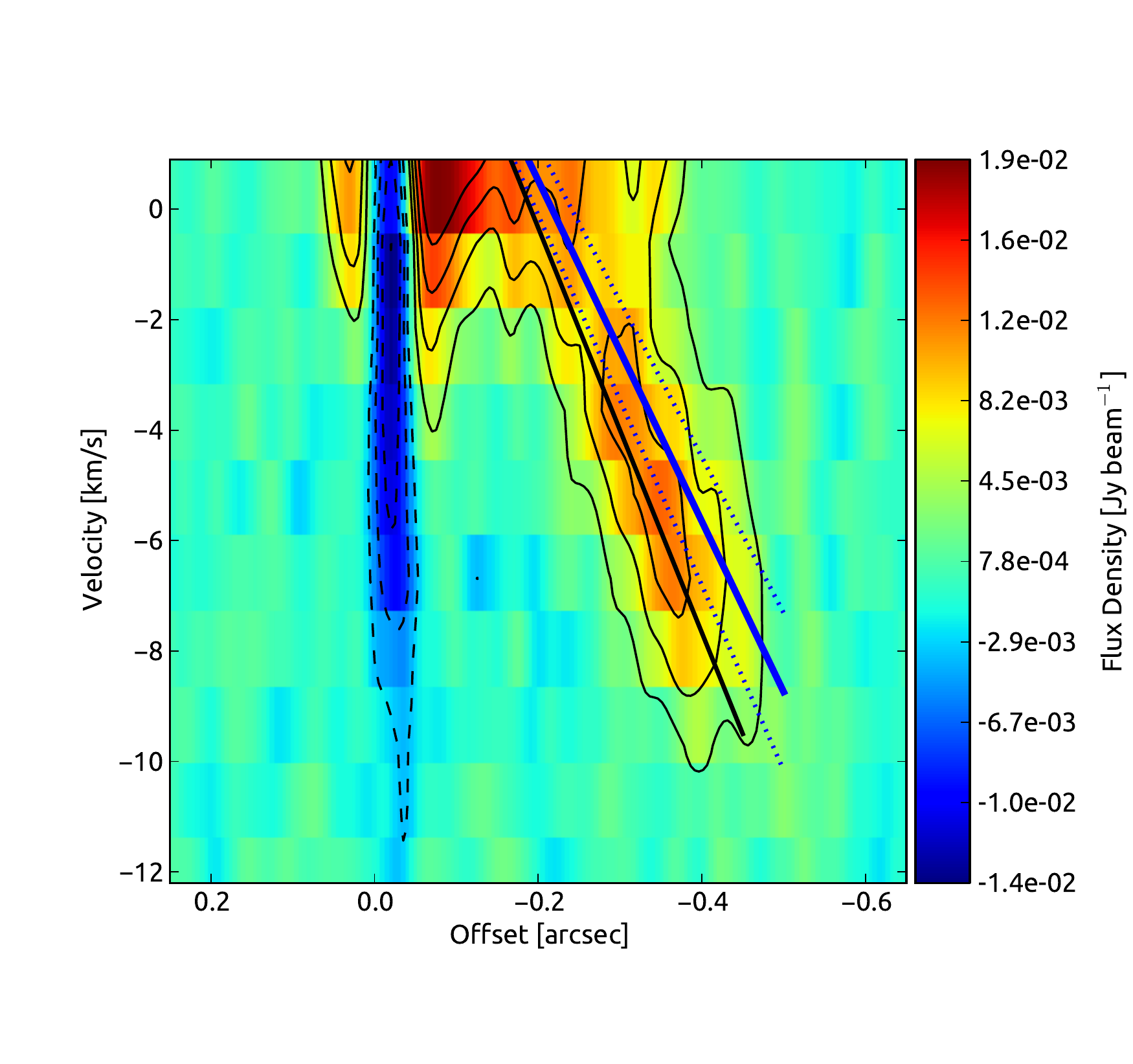}
\caption{Position velocity diagram of the $^{29}$SiO $v=0, J=5-4$ line
 at a rest frequency of $214.385741$~GHz. The diagram is extracted
 along a slit through the star with a width of $0.1\arcsec$ at a
 position angle of $135^\circ$. Contours are drawn at $-16$ to
 $16\sigma$ in steps of $4\sigma$, where
 $\sigma=0.9$~mJy~beam$^{-1}$. We only show the blue-shifted channels
 to highlight the linear feature extending to $\sim0.4\arcsec$. The
 offset is with respect to the star. The stellar velocity is
 $6.7$~km~s$^{-1}$. The blue lines indicates solid body rotation with
 an angular velocity of $(3.5\pm0.3)\times10^{-9}$~rad~s$^{-1}$ (for
 a position angle of the rotation axis of $45^{\circ}$, the dotted
 lines correspond to the quoted uncertainties). The black
 line corresponds to an angular velocity of
 $6.7\times10^{-9}$~rad~s$^{-1}$ when assuming a position angle of
 the rotation axis of $7^\circ$. The absorption feature against the
 star is due to the dynamic stellar atmosphere.}
  \label{PV}
\end{figure}

In addition to the lines emitting close to the star, we also detect a
feature extending out to $\sim0.4\arcsec$ (corresponding to
$>10$~R$_*$) at an anomalous velocity in a number of other lines. In
our observations it is most obvious in the $^{29}$SiO $v=0, J=5-4$
line for which we present a position-velocity diagram in
Fig.~\ref{PV}. As indicated in this diagram, the
velocity of this extended feature is consistent with the solid-body
rotation, determined in the other lines, out to a much larger
distance. However, the feature is seen in the south-east at a position
angle of $\sim135^\circ$ which would imply a position angle of the
rotation axis of $\sim45^\circ$. If the position angle instead is
$\sim7^\circ$ as determined from the other lines, the solid-body
angular velocity of the feature would be
$\sim6.7\times10^{-9}$~rad~s$^{-1}$. No corresponding feature is seen
on the other side of R~Dor.

\begin{table*}
\caption{Rotation model fit results}             
\label{tab1}      
\centering          
\begin{tabular}{l l c c c c c c c }     
\hline\hline       
Line  & Rest Freq. & $R_{\rm shell}$ & $\upsilon_{R, {\rm rot}} \sin{i}$ & $\omega_R \sin{i}$ & $\upsilon_{\rm sys}$ & $\upsilon_{\rm exp}$ &
$\Phi$ & $\chi^2_{\rm red}$\\
 & [GHz] &  [mas] & [km~s$^{-1}$] & [$10^{-9}$ rad~s$^{-1}$] & [km~s$^{-1}$]& [km~s$^{-1}$] &
 [$^\circ$] & \\
\hline
SiO $v=3, J=5-4$ & 212.5826 & $36\pm1$ & $-1.4\pm0.3$ & $4.7\pm0.8$ & $4.9\pm0.2$ & $1.7\pm0.2 $& $-1\pm11$ & 1.13 \\
SO$_2$ $J_{K_{\mathrm{a}},K_{\mathrm{c}}} = 16_{3,13}-16_{2,14}$ &
214.68939 & $66\pm2$ & $-1.8\pm0.1$ & $3.2\pm0.2$ & $6.7\pm0.1$ & $0.0\pm0.1$ & $9\pm4$ & 0.89 \\
$^{29}$SiO $v=1, J=5-4$$^a$ & 212.90515 & - & $-4.1\pm1.0$$^b$ & $15\pm4$ & $3.1\pm0.2$ & - &
$3\pm16$ & 1.46 \\
\hline                  
\end{tabular}
\tablefoot{
\tablefoottext{a}{Only seen in absorption against the star. The fit
 represents the brightest of two absorption components.}
\tablefoottext{b}{The velocity at $R_*$.}}
\end{table*}


\section{Discussion}

\subsection{The origin of the rotation}

Standard angular momentum conservation considerations imply that
TP-AGB stars should be slow rotators. The actual surface rotation is
affected by a variety of processes, such as mass loss and magnetic
breaking, during stellar evolution. We compare our observations with a
detailed model for a star with an initial mass of $1.5$~M$_\odot$ and
an initial rotation velocity of $50$~km~s$^{-1}$, which use the same
code and physics as explained in \citet{GS2014}. The velocity
corresponds to a typical magnetically braked main sequence rotation
velocity for such a star \citep[e.g.][]{Calvet1983}. The velocity at
which the centrifugal force balances gravity, in the absence of
radiation pressure, is $\sim250$~km~s$^{-1}$, but the majority of
stars rotate slower than $40\%$ of this velocity on the main sequence
\citep{Huang2010}.  Once the star reaches the thermally-pulsing AGB
phase, and reaches the measured temperature ($\sim2100$~K) and radius
($\approx 400$~$R_\odot$ at 59~pc), the surface rotation velocity has
dropped to $\sim0.01$~km~s$^{-1}$.  The rotation velocity remains of
this order even when higher initial main sequence rotational
velocities are considered \citep{GS2014}. This is at least two orders
of magnitude less than what we find. Unless the theory is
incomplete, it is unlikely that the rotation we measure originates in
the fast initial rotation of a single star.

In \citet{GS2016}, stellar rotation was determined in the presence of
a companion. In these models, it is possible to achieve velocities
similar to what we observe by invoking a companion within $8$~au. For
sub-stellar companions, we can also constrain the inner radius to be
$\gtrsim2.5$~au considering that during the pre-AGB evolution this region
would be gravitationally cleared \citep{Mustill2012}. For illustration purpose, a
$0.1$~M$_\odot$ companion at $5$~au on a circular orbit would have a
total angular momentum $\sim2\times10^{45}$~kg~m$^2$~s$^{-1}$. This
corresponds roughly to ten times the angular momentum estimated to be
imparted to R~Dor.

Thus, the most likely cause of the observed rotation is a yet
undetected low-mass companion. However, this does not explain the
apparent solid-body rotation out to several stellar
radii. Specifically, the origin of the fast rotating feature out to
$>10$~R$_*$ remains unclear. If this feature is somehow related to the
rotation one might expect it to be found close to the equatorial
plane. As the position angle of the feature is very different from
that of the rotation axis, this would require a relatively small
inclination $i$ and hence an even larger rotation velocity, or a
significant change of the rotation axis. Alternatively, this feature
could be unrelated to the rotation and represent a seemingly one-sided
ejection of material.

\subsection{The effect of the rotation}
The critical velocity, for rotation having an influence on the AGB
atmospheric structure, is set by the local sound speed
\citep[e.g.][]{Ignace1996, GS2014}. For typical AGB atmospheres this
velocity ranges from $\sim5$~km~s$^{-1}$ near the stellar photosphere
to $\sim1$~km~s$^{-1}$ further out in the circumstellar envelope
(CSE). Our measured velocity at $\sim2~R_*$ is similar to the local
sound speed, assuming typical temperature and density, and would even
exceed it if $i\lesssim 40^\circ$. The rotation is thus expected to
have a measurable effect on the density distribution through the
CSE. Specifically, models predict a density contrast between the equatorial and polar regions
  exceeding a factor of three \citep{Ignace1996, Dorfi1996}.
No strong
effect on the circumstellar environment of R~Dor are immediately
apparent. This could be consistent with a rotation axis inclination
angle $\gtrsim 45^\circ$, which would imply we mainly see the
equatorial region projected on the plane of the sky. Finally, we note
that our measured angular rotation could be sufficient to sustain a
rotational dynamo although the magnetic field strength of R~Dor is yet
to be determined.

\section{Conclusions}

We present the first direct observations of rotation of an AGB star
other than the fast rotating star V~Hya, which 
is thought to be transitioning to the
post-AGB stage \citep[e.g.][]{Sahai2016}. We measure an apparent angular rotation velocity of
the oxygen-rich AGB star R~Dor of
$\omega_R\sin{i}=3.5\pm0.3\times10^{-9}$~rad~s$^{-1}$, equivalent
to a rotation velocity of $|\upsilon_{\rm rot}\sin{i}|=1.0\pm0.1$~km~s$^{-1}$ at
the observed (214~GHz) stellar radius of $R_*=1.84$~au. The solid-body
rotation apears to extend to at least 2~$R_*$, where we measure a
rotation velocity $|\upsilon_{\rm
  rot}\sin{i}|\sim1.8\pm0.1$~km~s$^{-1}$. 

Although the inclination of the rotation axis is unknown, the rotation
velocity exceeds what can be produced by single star evolution models
by almost two orders of magnitude. This would suggest that R~Dor has
an unseen companion from which angular momentum is transferred to its
extended atmosphere. No such companion has been directly
detected. A study of the dust scattered light in the dust formation
zone of R~Dor also does not show obvious asymmetries due to an
interaction \citep[][and in prep.]{Khouri2016}. Furthermore, we
are also not aware of any reported large scale bipolar outflow or
spiral structure that could be the result from binary
interaction. \citet{Schoeier2004} do note a $\sim1\arcsec$-scale
asymmetry in the SiO $v=0, J=2-1$ envelope.

Since R~Dor is the nearest known AGB star, it should be possible
  to identify the nature of the companion. It will then be a prime
candidate to observe close binary interaction in detail, and
specifically investigate binary and/or rotation effects on dust
formation and the generation of AGB winds. As the rotation velocity
approaches the local sound speed, the mass loss should no longer be
fully isotropic. Hence R~Dor might also be a progenitor of an
aspherical planetary nebula. 

\begin{acknowledgements}
  This work was supported by ERC consolidator grant 614264. WV, TK and
  HO acknowledge support from the Swedish Research Council. EDB
  acknowledges support from the Swedish National Space
  Board. This paper makes use of the following ALMA data:
  ADS/JAO.ALMA\#2016.1.00004.S. ALMA is a partnership of ESO
  (representing its member states), NSF (USA) and NINS (Japan),
  together with NRC (Canada), NSC and ASIAA (Taiwan), and KASI
  (Republic of Korea), in cooperation with the Republic of Chile. The
  Joint ALMA Observatory is operated by ESO, AUI/NRAO and NAOJ.
\end{acknowledgements}

\bibliographystyle{aa}
\bibliography{RDorrefs}

\begin{appendix}

\section{Supplementary material}
\label{maps}

Here we present the spectra for the SiO $v=3, J=5-4$,
SO$_2$ $J_{K_{\mathrm{a}},K_{\mathrm{c}}} = 16_{3,13}-16_{2,14}$, and $^{29}$SiO $v=1,
J=5-4$ lines. We also show the velocity map and model for the  $^{29}$SiO $v=1,
J=5-4$ line as well as the error, residual, intensity and fwhm
velocity maps produced by {\it specfit} for the SiO $v=3, J=5-4$,
SO$_2$ $J_{K_{\mathrm{a}},K_{\mathrm{c}}} = 16_{3,13}-16_{2,14}$, and $^{29}$SiO $v=1,
J=5-4$ lines.

 \begin{figure*}
\centering
\includegraphics[width=5cm]{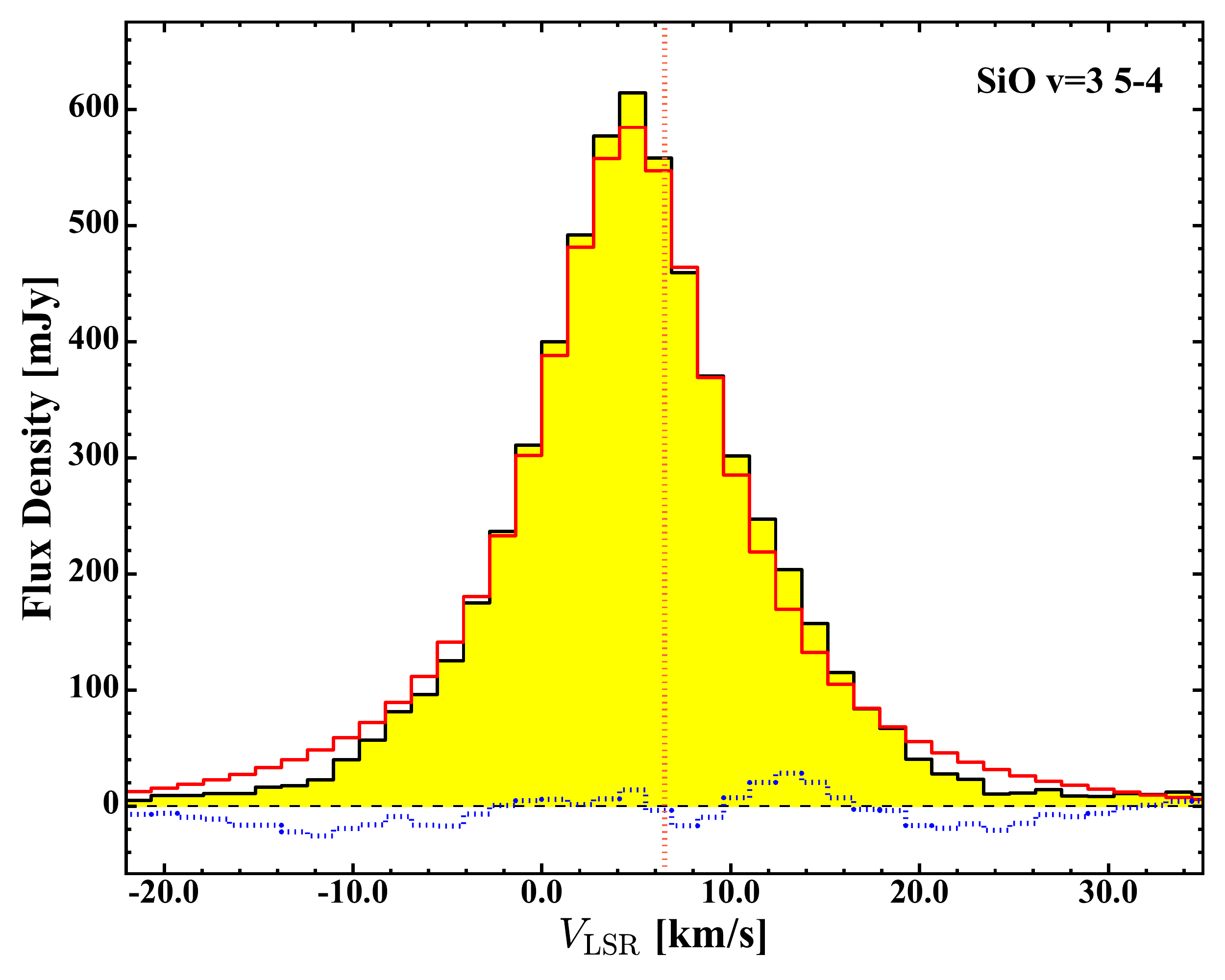}
\includegraphics[width=5cm]{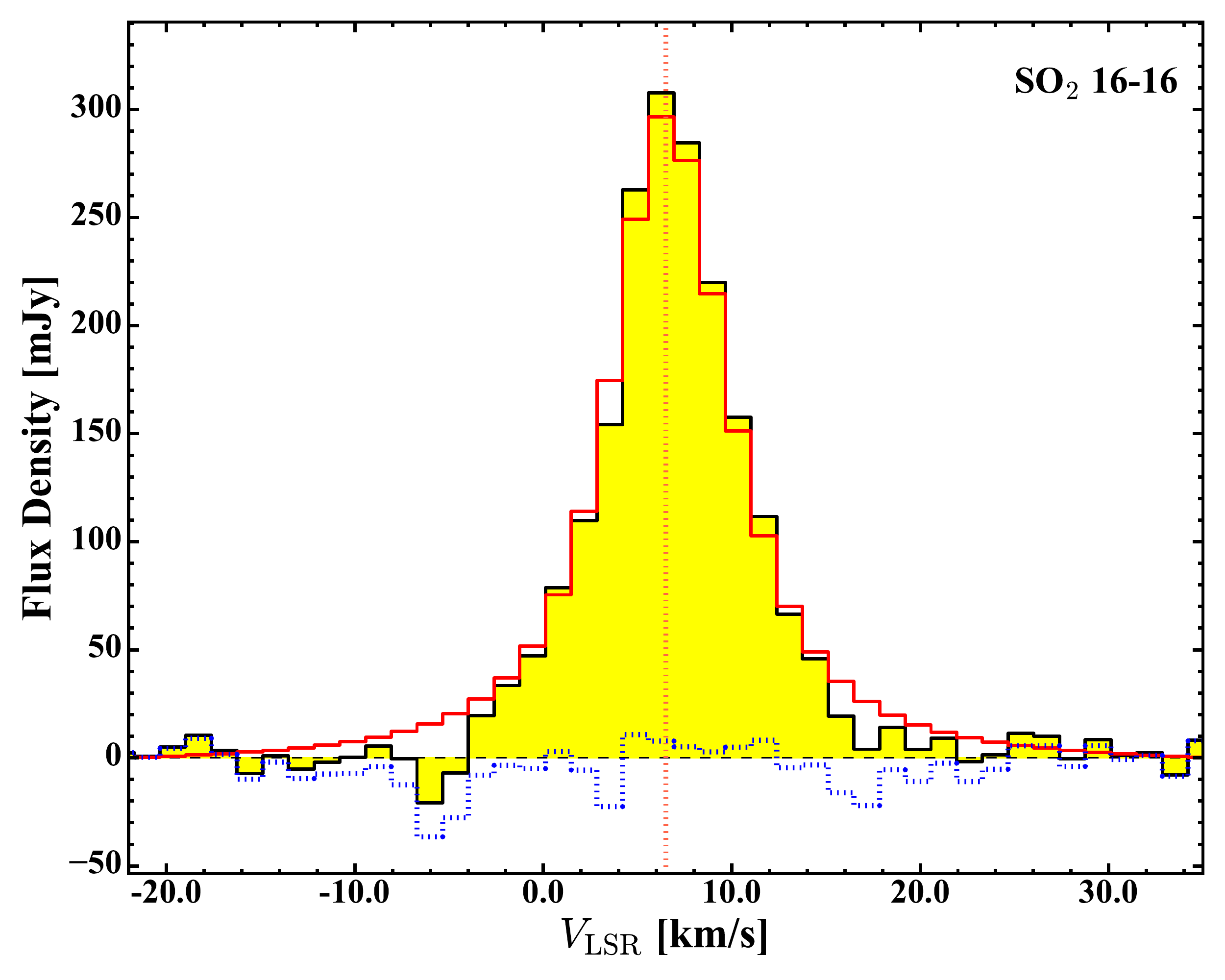}
\includegraphics[width=5cm]{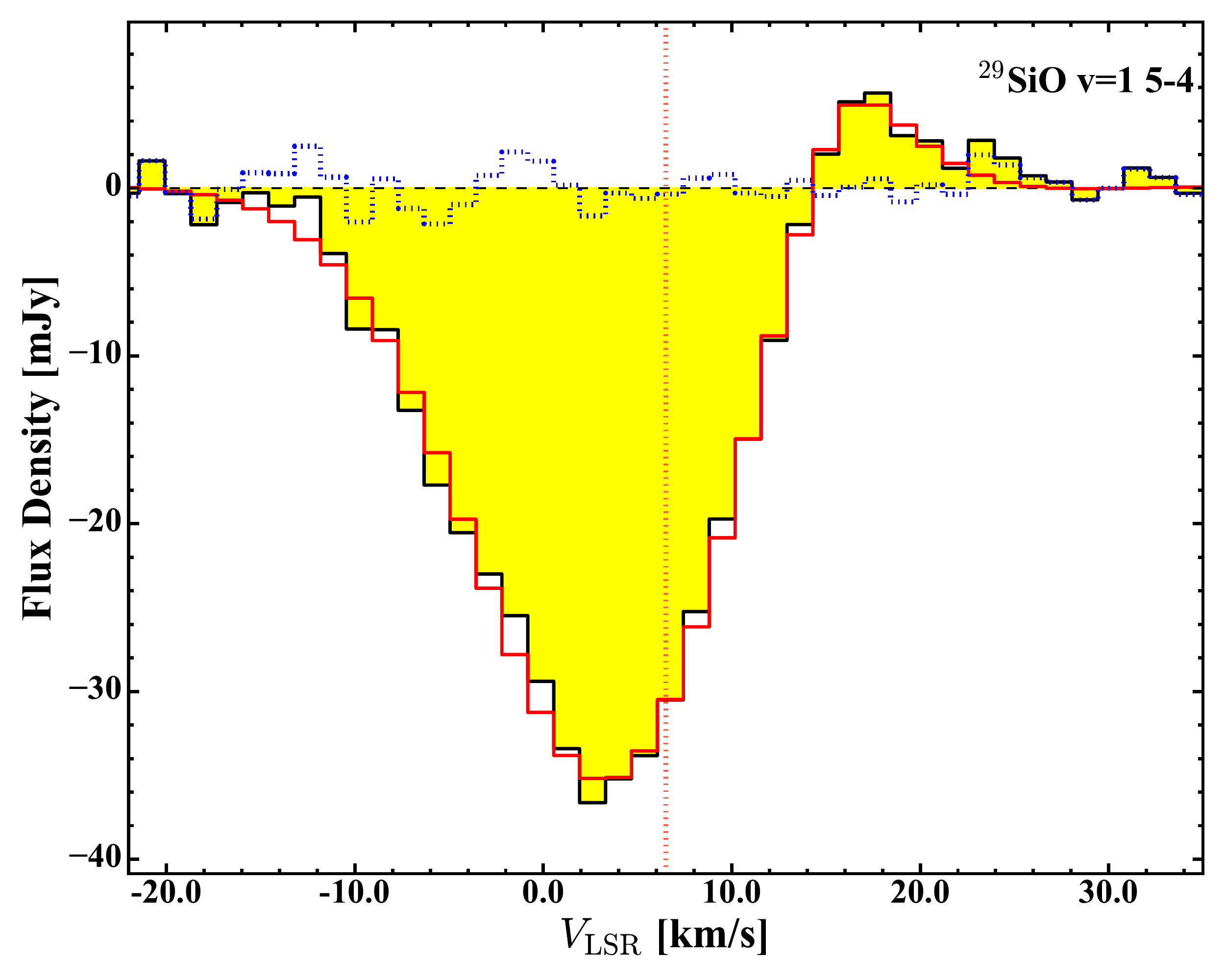}
\caption{Total intensity spectrum of the $^{28}$SiO $v=3 J=5-4$ {\it
   (left)} and the SO$_2$ $J_{K_{\mathrm{a}},K_{\mathrm{c}}} =
 16_{3,13}-16_{2,14}$ {\it (middle)}, and the $^{29}$SiO $v=1, J=5-4$
 {\it (right)} transitions around R~Dor. The spectra
 are extracted using a circular aperture with a diameter of
 $0.09\arcsec$, 
 $0.2\arcsec$, and $0.075\arcsec$ 
for the SiO $v=3$, SO$_2$, and $^{29}$SiO $v=1$ lines
respectively. The observed spectra are shown in
 black. The best-fit CASA {\it specfit} model for each line is shown
 in red, while the residuals are indicated by the dashed blue
 histogram. }
   \label{lines}
\end{figure*}

\begin{figure*}
\centering
\includegraphics[width=15cm]{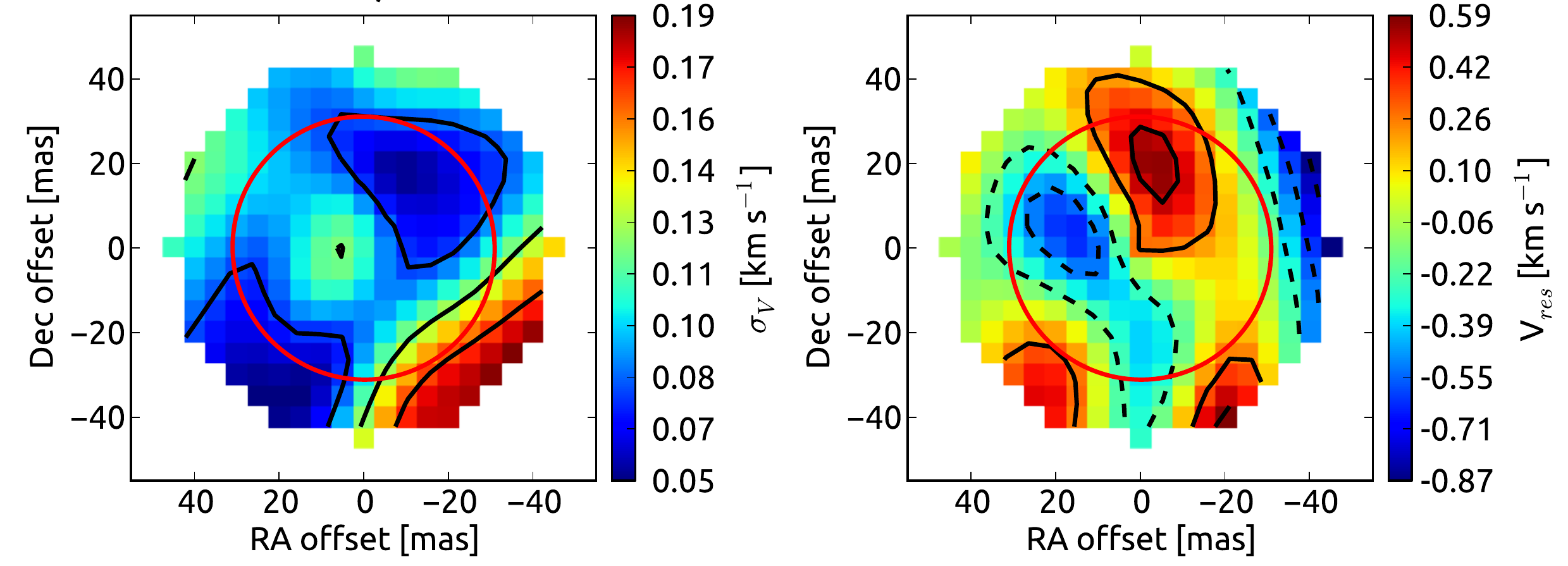}
\caption{{\it (left)}: The error map of
 velocity, as derived from the pixel-based spectral fit of the SiO
 $v=3, J=5-4$ line, used in the
 $\chi^2$ analysis. {\it (right)}: The velocity residual map.}
  \label{SiOres}
\end{figure*}

\begin{figure*}
\centering
\includegraphics[width=15cm]{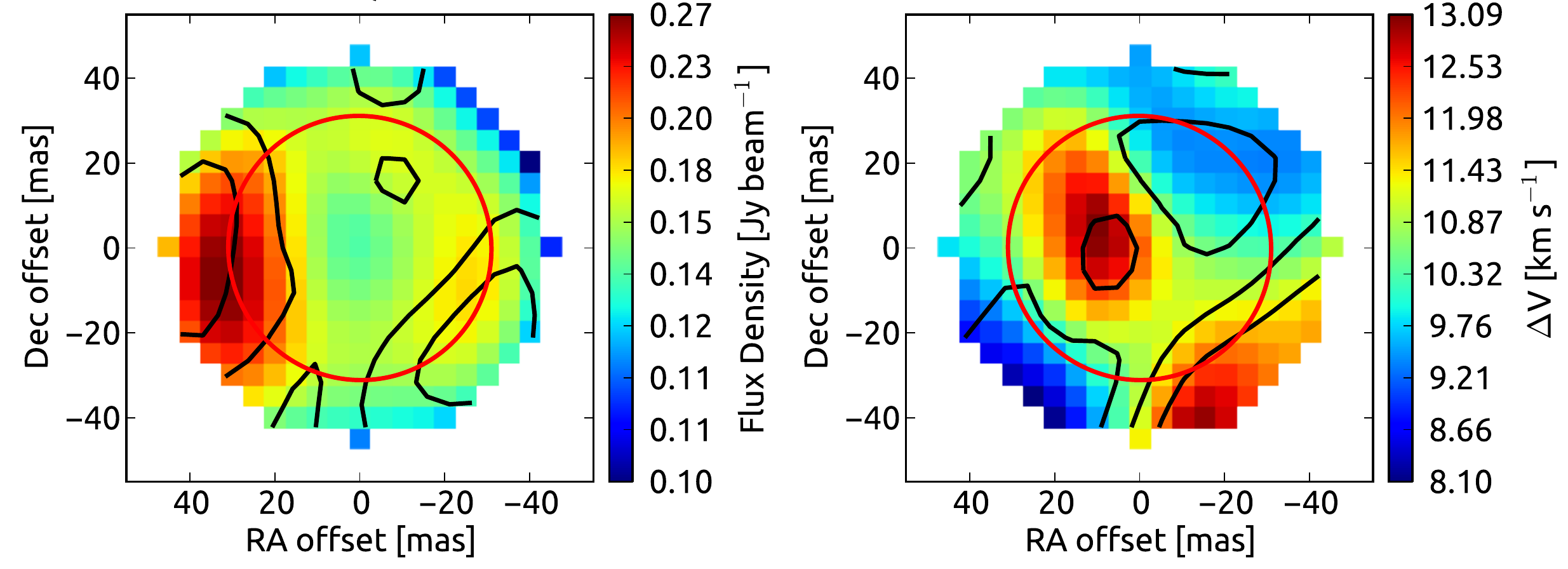}
\caption{{\it (left)}: The intensity map of SiO $v=3, J=5-4$
  as derived from the pixel-based spectral fit. The contours indicate
  the fitted rms errors at $2, 3,$~and~$4$~mJy~beam$^{-1}$. {\it (right)}: The fwhm velocity
  map. The contours indicate the fitted rms errors at levels of $0.26,
  0.39,$ and $0.52$~km~s$^{-1}$.}
  \label{SiOother}
\end{figure*}

\begin{figure*}
\centering
\includegraphics[width=15cm]{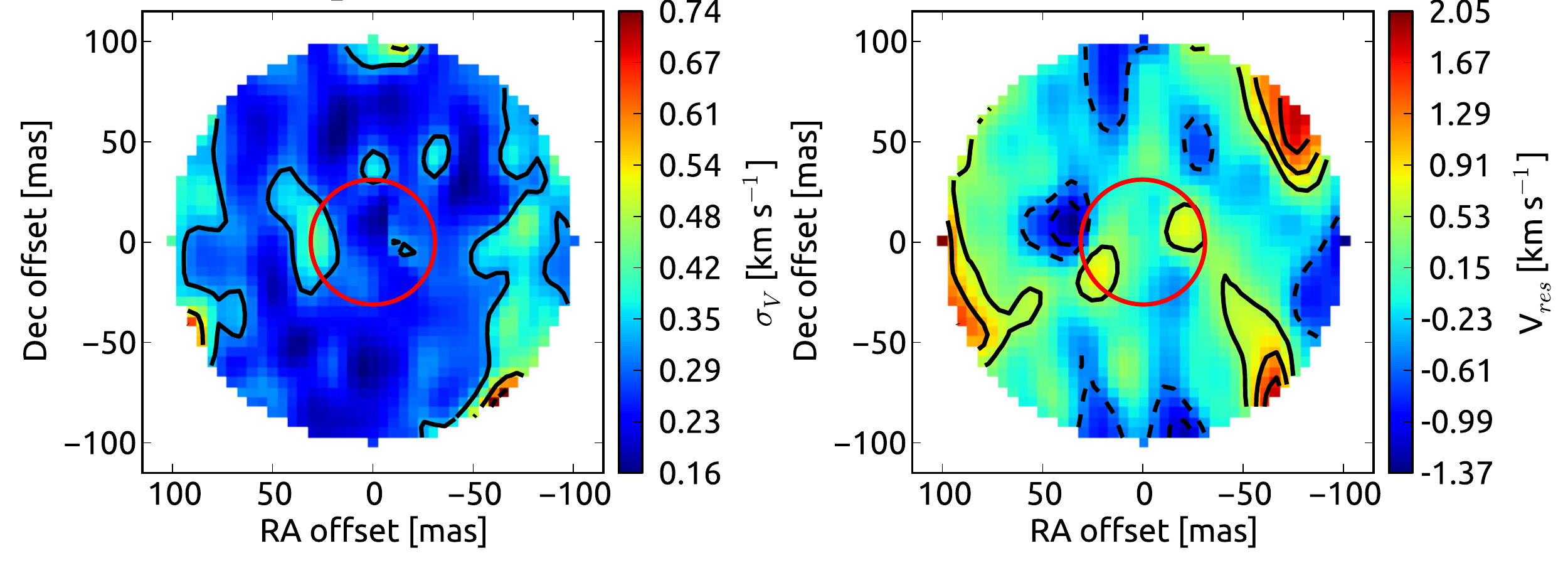}
\caption{As Fig.\ref{SiOres} for the SO$_2$ $J_{K_{\mathrm{a}},K_{\mathrm{c}}} = 16_{3,13}-16_{2,14}$ line.}
  \label{SO2res}
\end{figure*}

\begin{figure*}
\centering
\includegraphics[width=15cm]{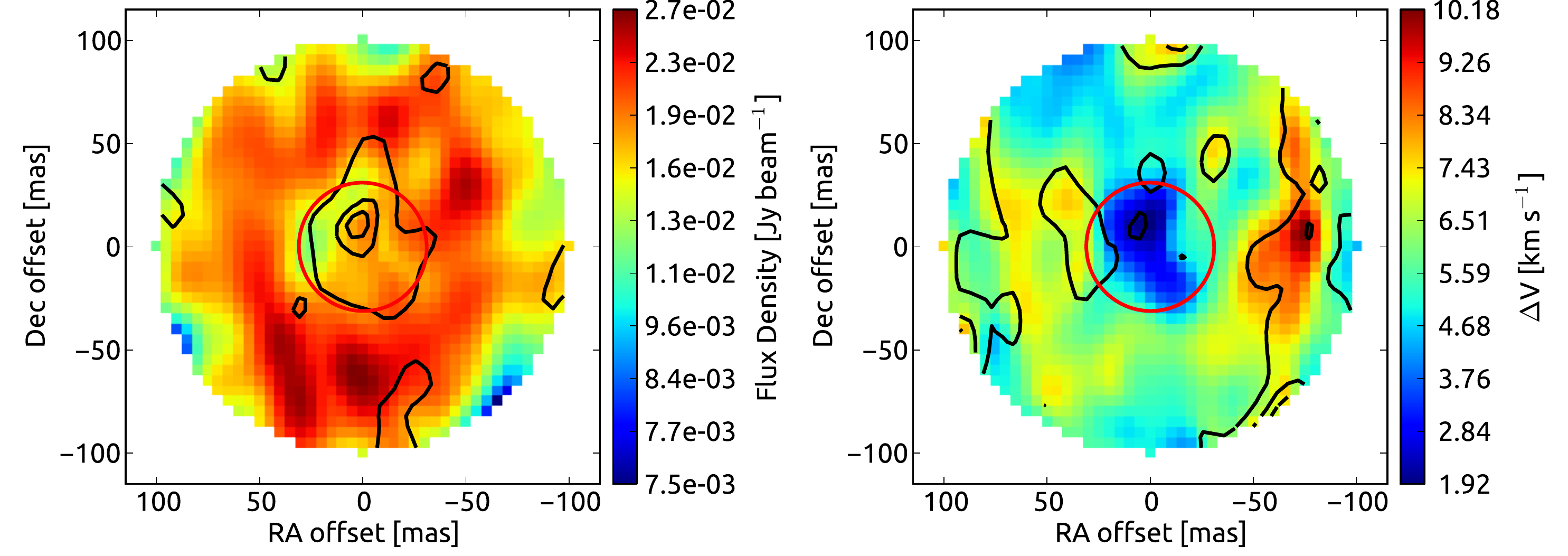}
\caption{As Fig.\ref{SiOother} for the SO$_2$
  $J_{K_{\mathrm{a}},K_{\mathrm{c}}} = 16_{3,13}-16_{2,14}$ line.  The
  fitted intensity errors are plotted at $2,
  4,$~and~$6$~mJy~beam$^{-1}$. The fitted fwhm velocity errors are
  plotted at $1.0,  1.5,$ and $2.0$~km~s$^{-1}$.}
  \label{SO2other}
\end{figure*}

\begin{figure*}
\centering
\includegraphics[width=14cm]{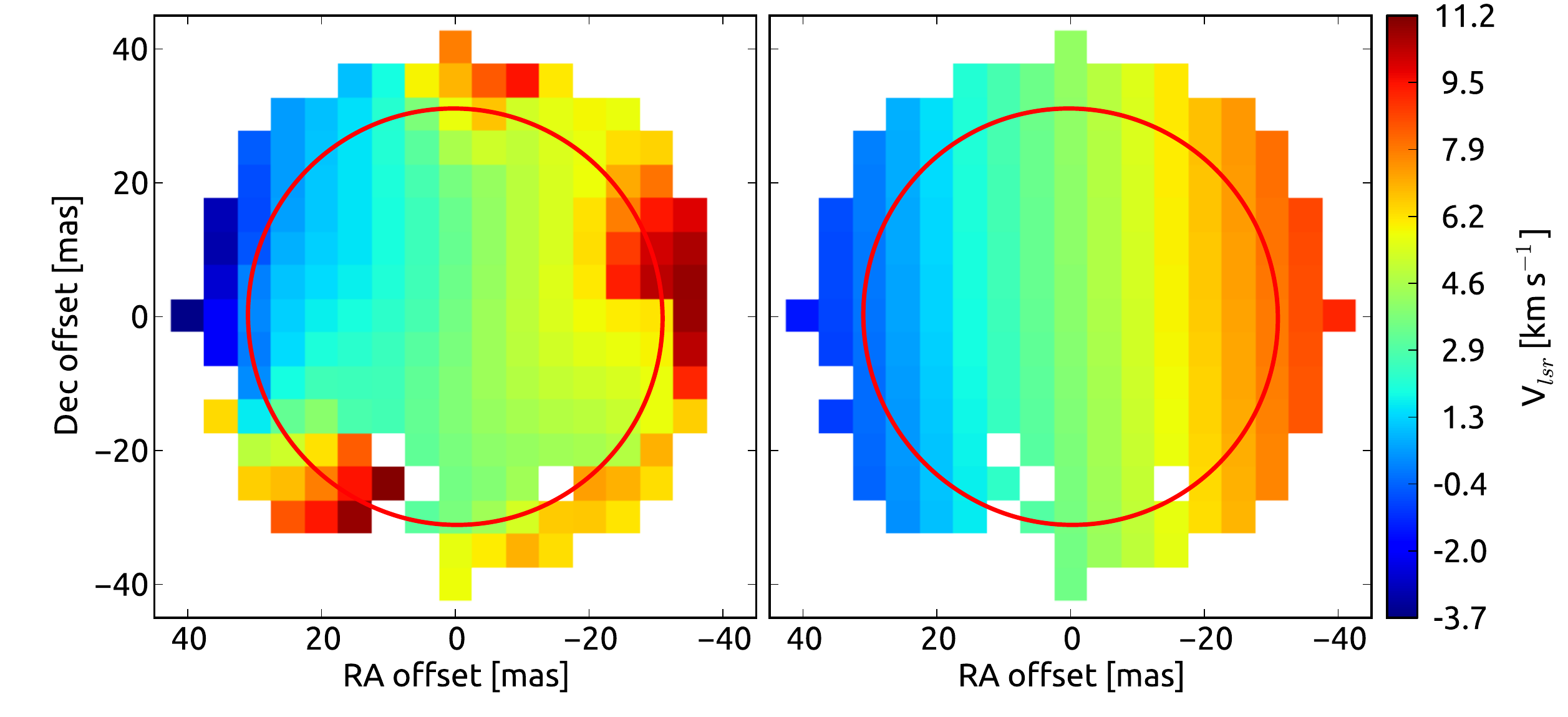}
\caption{As Fig.~\ref{SiOfit} for the dominant absorption component of
the $^{29}$SiO $v=1, J=5-4$ line.}
\label{29SiOfit}
\end{figure*}

\begin{figure*}
\centering
\includegraphics[width=7cm]{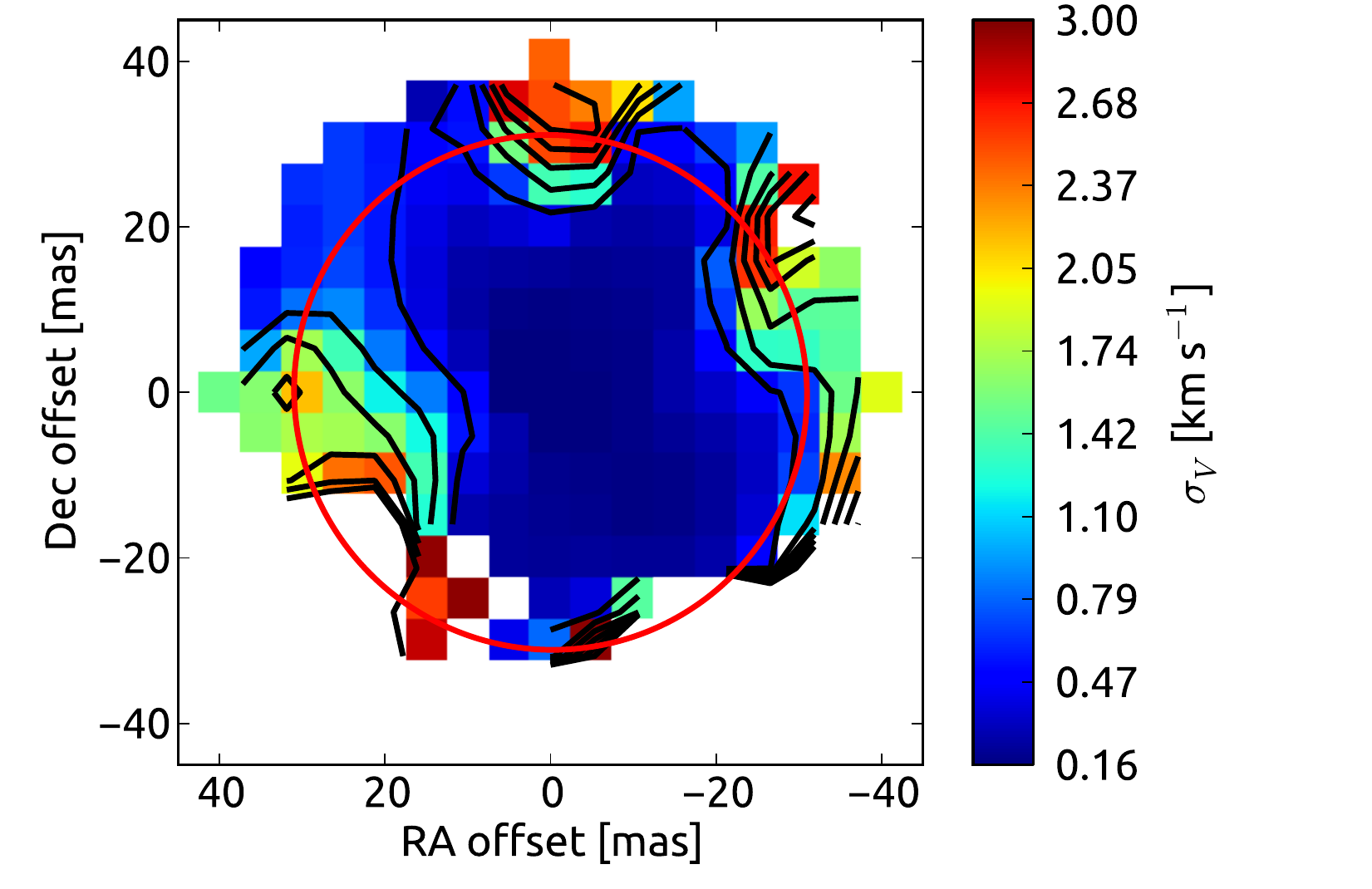}
\includegraphics[width=7cm]{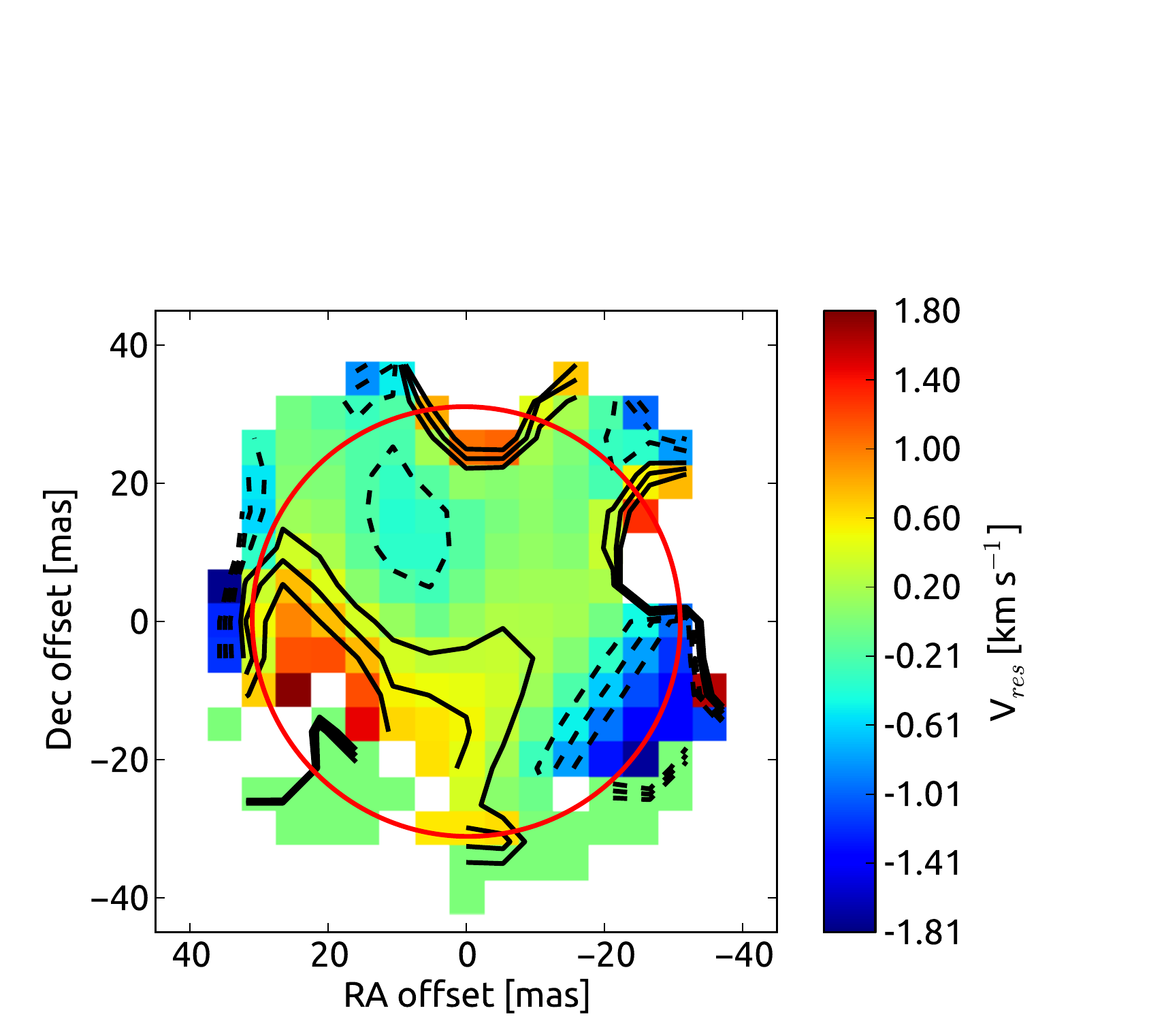}
\caption{As Fig.~\ref{SiOres} for the dominant absorption component of
the $^{29}$SiO $v=1, J=5-4$ line.}
\label{29SiOres}
\end{figure*}

\begin{figure*}
\centering
\includegraphics[width=7cm]{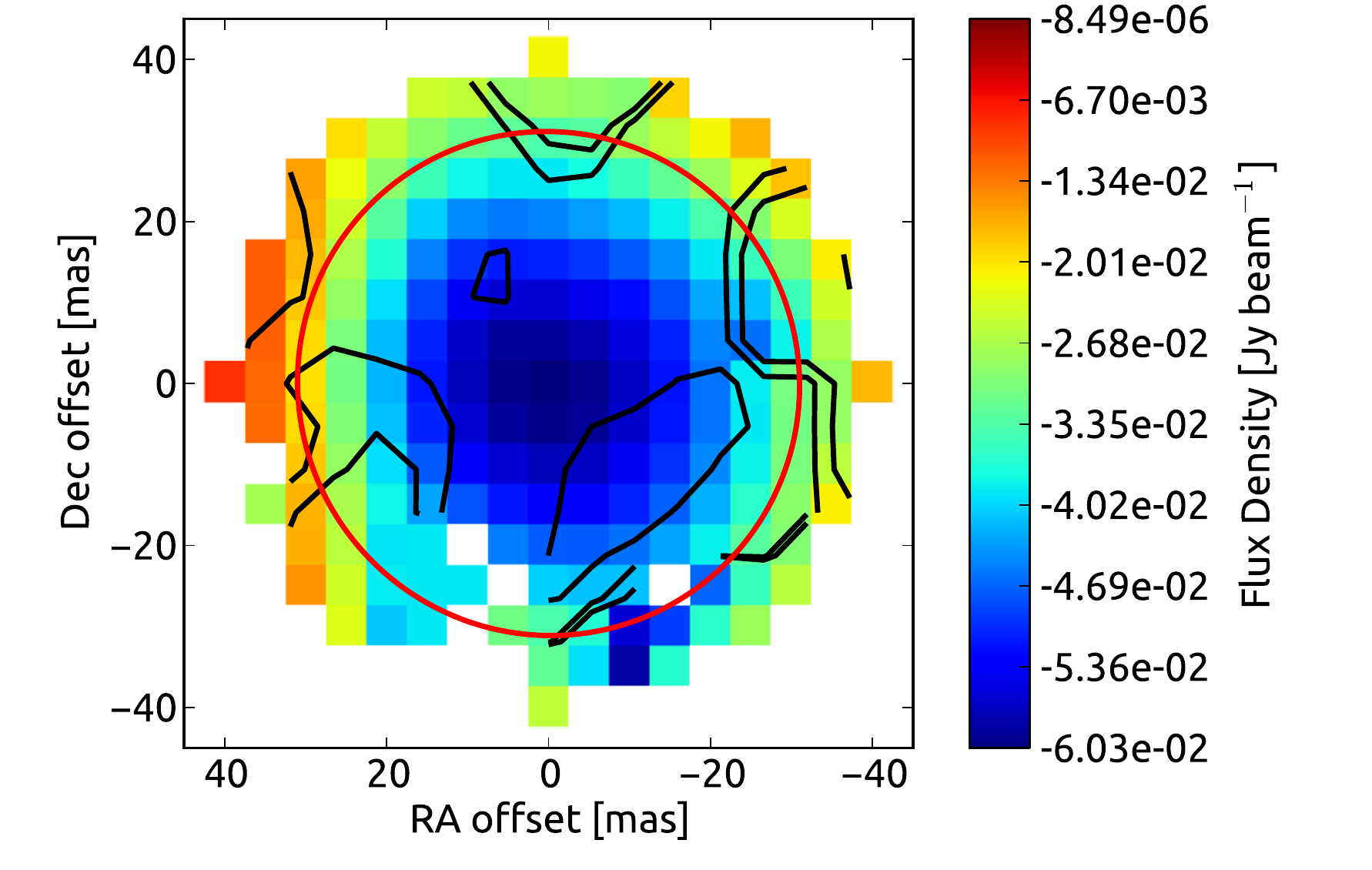}
\includegraphics[width=7cm]{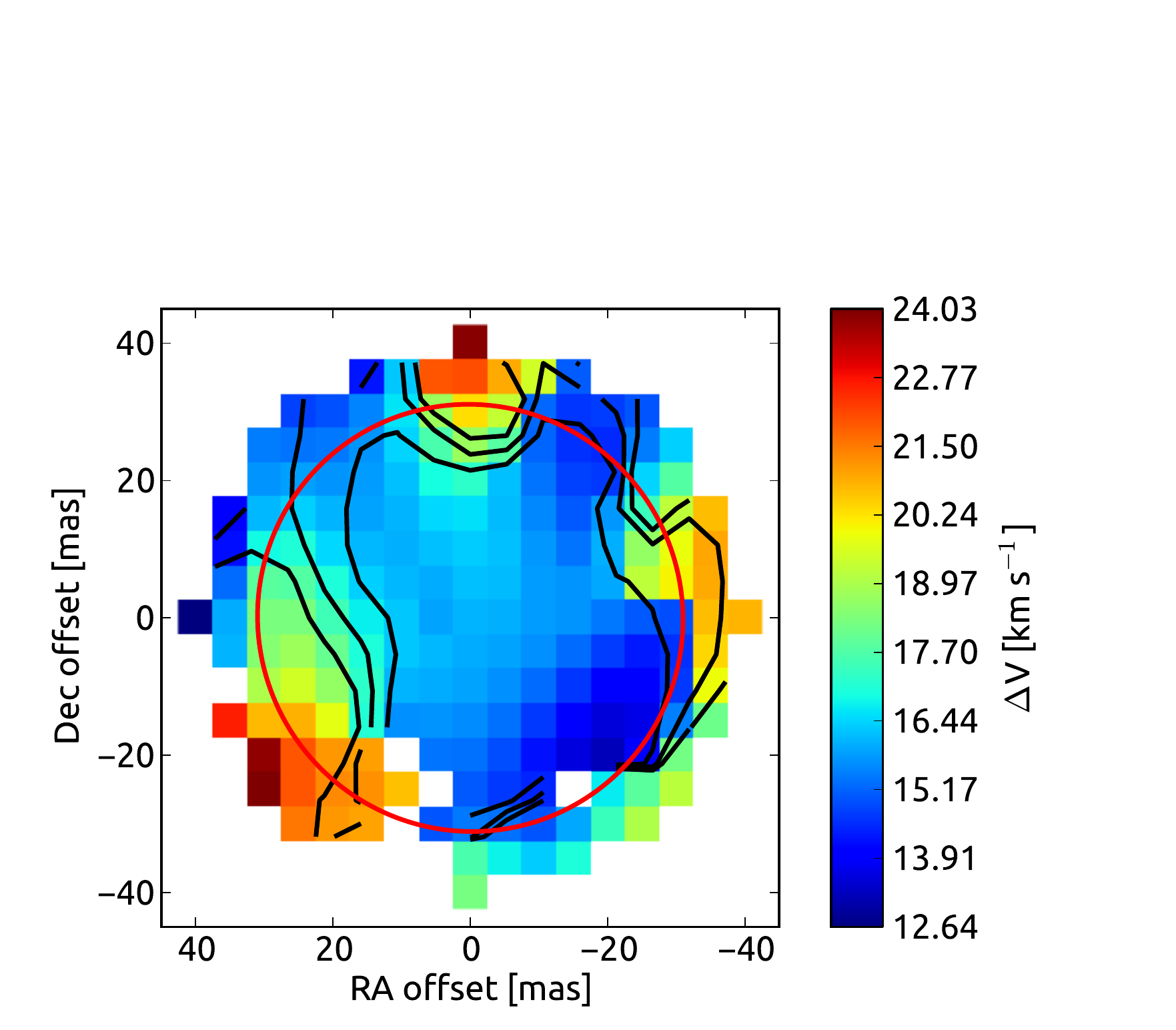}
\caption{As Fig.~\ref{SiOother} for the dominant absorption component of
the $^{29}$SiO $v=1, J=5-4$ line.  The
  fitted intensity errors are plotted at $1,
  2,$~and~$4$~mJy~beam$^{-1}$. The fitted fwhm velocity errors are
  plotted at $1.0,  1.5,$ and $2.0$~km~s$^{-1}$.}
\label{29SiOother}
\end{figure*}

\section{Channel maps}
\label{channels}
 
Here we present the channel maps for the SiO $v=3, J=5-4$,
SO$_2$ $J_{K_{\mathrm{a}},K_{\mathrm{c}}} = 16_{3,13}-16_{2,14}$, and $^{29}$SiO $v=1,
J=5-4$ lines.
\begin{figure*}
\centering
\includegraphics[width=15cm]{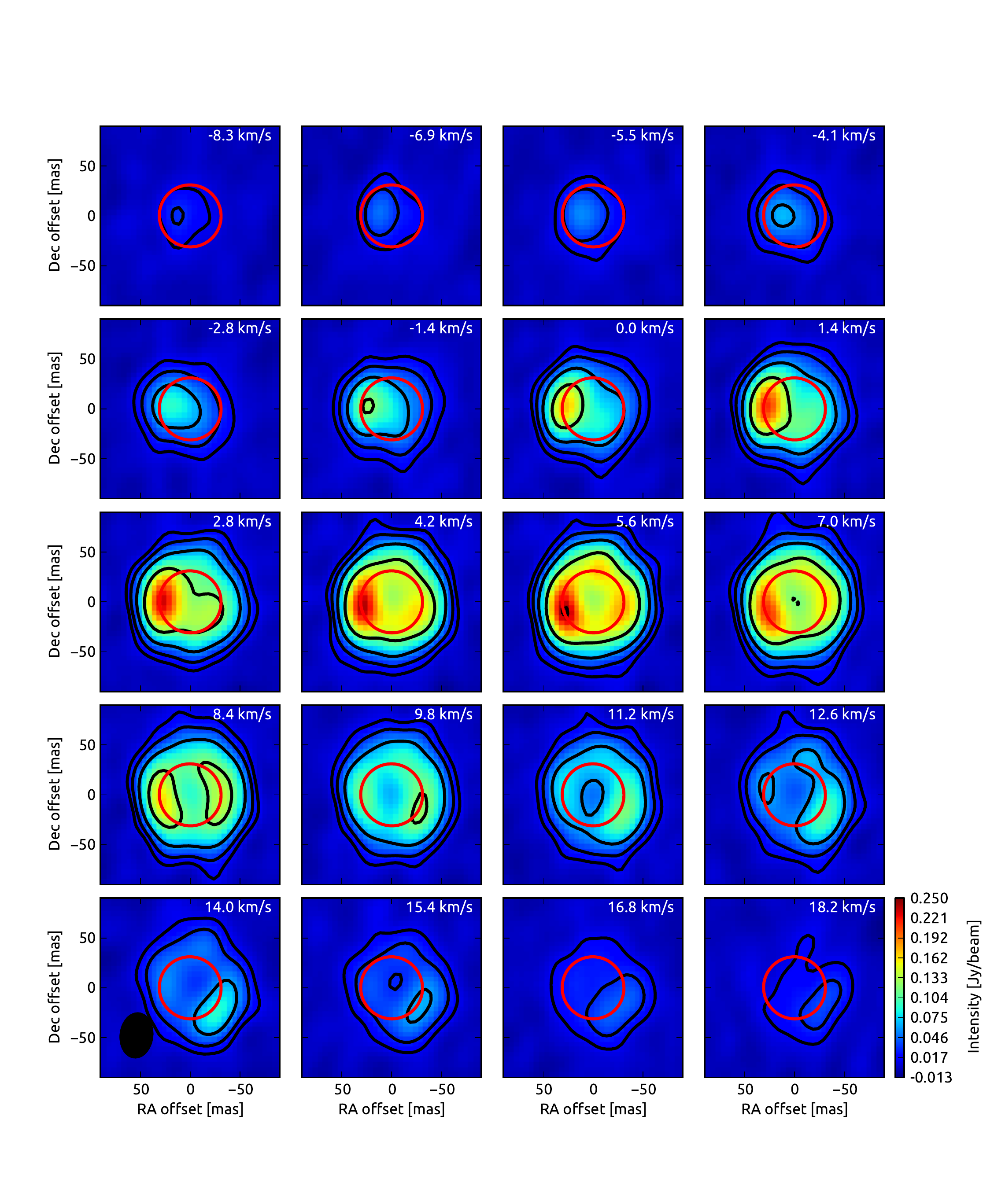}
\caption{Channel maps for the SiO $v=3, J=5-4$ emission line. The red
  ellipse denotes the size of the measured stellar disc. Contours are
  drawn at $6, 12, 24, 48, 96\sigma$ with $\sigma=2.5$~mJy~beam$^{-1}$. The beam size
  is indicated in the bottom left panel.}
\label{SiOv3maps}
\end{figure*}

\begin{figure*}
\centering
\includegraphics[width=15cm]{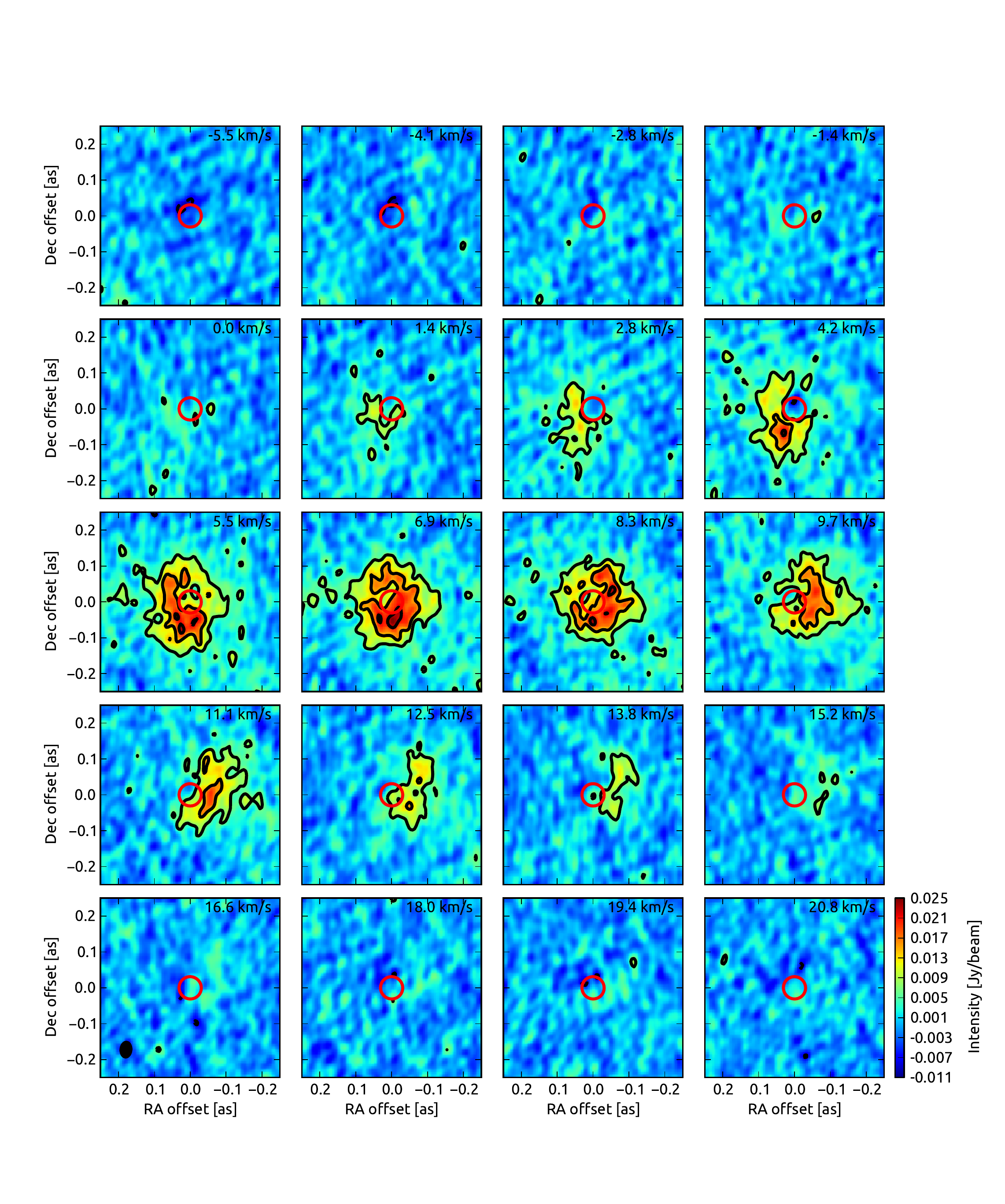}
\caption{As Fig.~\ref{SiOv3maps} for the SO$_2$
  $J_{K_{\mathrm{a}},K_{\mathrm{c}}} = 16_{3,13}-16_{2,14}$ emission
  line. Contours are
  drawn at $3, 6, 9, 12 \sigma$ with $\sigma=2.5$~mJy~beam$^{-1}$.}
\label{SO2maps}
\end{figure*}

\begin{figure*}
\centering
\includegraphics[width=15cm]{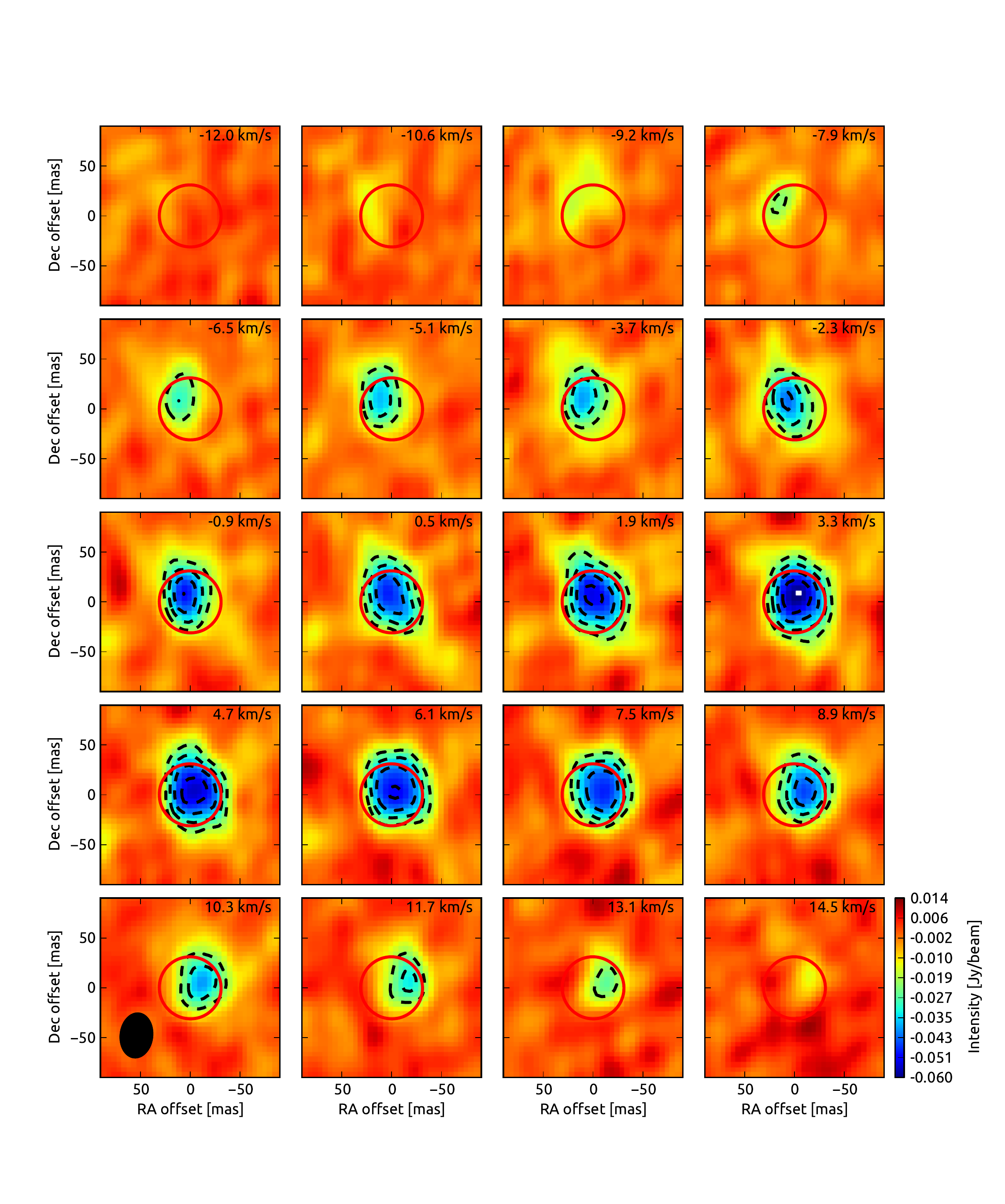}
\caption{As Fig.~\ref{SiOv3maps} for the $^{29}$SiO $v=1, J=5-4$ line
  seen in absorption. Contours are
  drawn at $-16, -12, -8\sigma$ with $\sigma=2.5$~mJy~beam$^{-1}$.}
\label{29SiOmaps}
\end{figure*}

\end{appendix}

\end{document}